\documentclass[aps,prd,twocolumn,groupedaddress,amsfonts,amssymb,
amsmath,showpacs,floatfix]{revtex4-1}

\usepackage{dcolumn}
\usepackage{multirow}
\usepackage{verbatim}
\usepackage{graphicx}
\usepackage{color}

\newcommand{\epstree}{\epsilon_{\text{tree}}}
\newcommand{\mtree}{m_{\text{tree}}}
\newcommand{\mkin}{m_{\text{kin}}}

\newcommand{\sigmaA}{\Sigma^{(p\!\!\!/)}}
\newcommand{\sigmaB}{\Sigma^{(\mathbb{I})}}

\newcommand{\python}{\texttt{{\normalsize P}{\footnotesize YTHON}}~}
\newcommand{\fortran}{\texttt{{\normalsize F}{\footnotesize ORTRAN}}~}
\newcommand{\vegas}{\texttt{{\normalsize V}{\footnotesize EGAS}}~}

\newcommand{\hippy}{\texttt{{\normalsize H}{\footnotesize I}{\normalsize
PP}{\footnotesize Y}}~}
\newcommand{\hpsrc}{\texttt{{\normalsize HP}{\footnotesize SRC}}~}
\newcommand{\mathematica}{\texttt{Mathematica}~}

\newcommand{\hippydot}{\texttt{{\normalsize H}{\footnotesize I}{\normalsize
PP}{\footnotesize Y}}.~}
\newcommand{\hippycomma}{\texttt{{\normalsize
H}{\footnotesize I}{\normalsize PP}{\footnotesize Y}},~}

\DeclareMathOperator*{\res}{\text{Res}}
\newcommand{\als}{\alpha_s}

\newcommand{\calf}{{\cal F}}
\newcommand{\cali}{{\cal I}}
\newcommand{\be}{\begin{equation}}
\newcommand{\ee}{\end{equation}}
\newcommand{\nl}{\nonumber \\}
\newcommand{\meff}{m_{\text{eff}}}

\newcommand{\intk}{\int\!\frac{\mathrm{d}^4k}{(2 \pi)^4}\;}

\begin{document}


\title{Matching lattice and continuum axial-vector and vector currents with
NRQCD and HISQ quarks}


\author{Christopher \surname{Monahan}}
\affiliation{Physics Department, College of William and Mary,
Williamsburg, Virginia 23187, USA}
\author{Junko \surname{Shigemitsu}}
\affiliation{Physics Department, The Ohio State University, Columbus,
Ohio 43210, USA}
\author{Ron \surname{Horgan}}
\affiliation{DAMTP, Centre for Mathematical Sciences, University of
Cambridge, Cambridge, CB3 0WA, UK}

\collaboration{HPQCD Collaboration}
\noaffiliation


\begin{abstract}
We match the continuum and lattice axial-vector and vector
currents at one loop in perturbation theory. For the heavy quarks we use
the nonrelativistic QCD (NRQCD) action and for the light quarks the Highly
Improved Staggered Quark (HISQ) action. We present results for both
massless and massive HISQ quarks and as part of the matching
procedure we include a discussion of the one loop HISQ renormalisation
parameters.
\end{abstract}

\pacs{12.38.Bx,12.38.Gc,13.20.Gd,13.20.He}

\maketitle



\section{Introduction}

Electroweak processes are an important tool in understanding the Standard
Model (SM) of particle physics, serving as an input into tests of the
unitarity of the Cabibbo-Kobayashi-Maskawa (CKM) matrix
and as a probe for new physics. The hadronic matrix elements that
characterise the strong interaction dynamics of these processes are a
crucial ingredient in the determination of CKM unitarity.

Global fits to the CKM unitarity have, in recent years,
indicated some tensions at the 2-3$\sigma$ level within the SM
\cite{lunghi11a,lunghi11b,laiho10,laiho12}. In many cases, the constraints
on the CKM unitarity triangle are limited by the precision with which the
nonperturbative inputs are known and thus it is imperative that these
inputs are determined as precisely as possible.

The
HPQCD collaboration has undertaken a suite of precision calculations of
heavy-light mesons as part of a program to precisely determine nonperturbative
contributions to electroweak parameters. Recent calculations of the decay constants $f_B$ and
$f_{B_s}$ have achieved a precision at the $2\%$ level, by taking advantage 
of the small discretisation errors and good chiral properties of the Highly Improved Staggered Quark
(HISQ) action \cite{mcneile12a,na12}. These results represent the most precise
currently available for these decay constants. In addition, nonperturbative
studies of the heavy-light semileptonic decays $B\rightarrow \pi\ell\nu$,
$B\rightarrow K\ell^+\ell^-$ and $B_s\rightarrow K\ell\nu$ are underway
\cite{bouchard12}.

The work of Ref.~\cite{na12} and \cite{bouchard12} use HISQ light quarks and
the
nonrelativistic QCD (NRQCD) action for the heavy quarks. These
calculations require matching the heavy-light axial-vector and vector
currents in the effective theory on the lattice with full QCD. In this
article we report on the one loop perturbative matching of the HISQ-NRQCD
axial-vector and vector current matching for both massless and massive
HISQ quarks. As part of
this procedure we determine the mass and
wavefunction renormalisation for massive HISQ quarks. Our matching results for
massive HISQ quarks will be relevant for future
studies of heavy-heavy decays $B_{(s)}\rightarrow D_{(s)}\ell\nu$.

In the next section we describe the quark and gluon actions used in
our calculation. We then review the formalism for extracting
renormalisation parameters from relativistic lattice actions and apply
these procedures to first massless and then massive HISQ quarks. We include
results for the one loop NRQCD mass and wavefunction renormalisation in
Section \ref{sec:nrqcdparms}. In Section \ref{sec:matching} we outline the
calculation of the matching coefficients and then, in Section
\ref{sec:matchresults}, we present our
results for a range of heavy quark masses. We conclude with a summary in
Section \ref{sec:summary}.

\section{The Lattice Actions}
\subsection{Gluon Action}

We use the Symanzik improved gluon action with tree level
coefficients \cite{weisz83,weisz84,curci83,luscher85a}, given by
\begin{equation}
S_G = -\frac{\beta}{3u_0^4} \sum_{x,\mu>\nu}\left[5P_{\mu\nu}
-\frac{1}{4u_0^2}\left(R_{\mu\nu}+R_{\nu\mu}\right)\right].
\end{equation}
Here $P_{\mu\nu}$ is the plaquette,
\begin{equation}
P_{\mu\nu} = \frac{1}{N_c}\text{Re}\text{Tr}\Big\{U_{\mu}(x)
U_{\nu}(x+\hat{\mu})U_{\mu}^{\dagger}(x+\hat{\nu})U_{\nu}^{\dagger}(x)
\Big\},
\end{equation}
and $R_{\mu\nu}$ the six-link loop,
\begin{align}
R_{\mu\nu} = {} & \frac{1}{N_c}\text{Re}\text{Tr}\Big\{U_{\mu}(x)
U_{\mu}(x+\hat{\mu})U_{\nu}(x+2\hat{\mu}) \nonumber \\
{} & \times U_{\mu}^{\dagger}(x+\hat{\mu}+\hat{\nu})
U_{\mu}^{\dagger}(x+\hat{\nu})U_{\nu}^{\dagger}(x)\Big\},
\end{align}
with $\beta = 2N_c/g^2$ and $u_0$ the tadpole improvement factor
\cite{lepage93}. Radiative improvements to the gluon
action do not contribute to the one loop matching calculation. In general, radiative
improvement generates an ${\cal O}(\alpha_s)$ insertion in the gluon
propagator. There are no external gluons in our calculation, so any such
improvements only contribute at two loops and higher.

We include a gauge-fixing term 
\begin{equation}
S_{\xi} = \frac{1}{2\xi}\sum_x\left[\sum_{\mu}\Delta_{\mu}
(aA_{\mu})\right]^2,
\end{equation}
where $\Delta_{\mu}$ is the symmetrised difference operator, which acts 
on the gauge fields as
\begin{equation}
\Delta_{\mu}A_{\mu}(x) \equiv A_{\mu}\left(x+\frac{\hat{\mu}}{2}\right) - 
A_{\mu}\left(x-\frac{\hat{\mu}}{2}\right),
\end{equation}
and $\xi$ is the gauge parameter. Where possible, we confirm that gauge invariant
quantities are independent of the choice of gauge parameter by working in
both Feynman, $\xi=1$, and Landau, $\xi=0$, gauges.

\subsection{Light Quark Action}

We discretise the light quarks in this work using the Highly Improved
Staggered Quark (HISQ) action \cite{follana07}. The HISQ action
significantly reduces taste breaking discretization errors and has been
used successfully to simulate both $b$ and $c$
quark systems \cite{follana08,mcneile10,gregory11,dowdall12b}. There are
two equivalent methods for writing staggered quark actions, using either
four component ``naive'' fermions or one component ``staggered'' fields
\cite{lepage99,wingate03}. Throughout this calculation we use the naive
fermion representation and we denote the bare quark mass $am_0$. In
Section \ref{sec:masslessres} we present our results for massless HISQ quarks,
corresponding to $am_0 = 0$. Before we present the quark actions used in
this work, we pause to briefly discuss some notation, which we summarise in
Table \ref{tab:msum}.
\begin{table}
\caption{\label{tab:msum} Summary of quark mass notation.
\\}
\begin{ruledtabular}
\begin{tabular}{cccc}
\vspace*{-5pt}\\
\multirow{4}{*}{HISQ} & $am_0$ & bare light quark mass \\
& $a\mtree$ & tree level pole mass \\
& $am_1$ & one loop pole mass \\
& $a\mkin$ & kinetic mass \\
\vspace*{-5pt}\\
\hline
\vspace*{-5pt}\\
NRQCD & $aM_0$ & bare heavy quark mass \\
\vspace*{-5pt}\\
\end{tabular}
\end{ruledtabular}
\end{table}
We use four
different quark mass definitions for relativistic HISQ quarks: the bare
quark mass; the tree level and one loop pole masses, $a\mtree$ and $am_1$
respectively; and the kinetic mass, $a\mkin$. We distinguish these
relativistic quark masses from the nonrelativistic quark mass in NRQCD by
using a lowercase $m$ for HISQ quarks and an uppercase $M$ for NRQCD
quarks. Only the bare heavy quark mass $aM_0$ is required for
nonrelativistic quarks in this calculation.

The starting point for constructing the HISQ action is the AsqTad action 
\cite{lepage99}, which is given by
\begin{equation}
S_{\text{AsqTad}} =
a^4\sum_x\overline{\psi}(x)\left(\gamma_{\mu}\nabla_{\mu}^{
\text{AsqTad}} + m_0\right)\psi(x),
\end{equation}
where the AsqTad operator is
\begin{equation}
\nabla_{\mu}^{\text{AsqTad}} = \nabla_{\mu}^F -
\frac{a^2}{6}(\nabla_{\mu})^3.
\end{equation}
Here the three-link term $(\nabla_{\mu})^3$ is referred to as the ``Naik''
term and the superscript $F$ indicates that we use fattened links in the
lattice difference operator $\nabla_{\mu}$. The fattened links are given
by
\begin{equation}
U_{\mu}(x) \rightarrow {\cal F}_{\mu}^{\text{AsqTad}}U_{\mu}(x),
\end{equation}
where
\begin{align}
{\cal F}_{\mu}^{\text{AsqTad}}= {} & 
\left[{\cal F}_{\mu}
-\sum_{\rho\neq\mu}\frac{a^2(\nabla_{\rho})^2}{4}\right], \label{eq:fasq} \\
{\cal F}_{\mu}= {} &
\prod_{\rho\neq\mu}\left(1+\frac{a^2\nabla_{\rho}^{(2)}}{4}\right)_{
\text{symmetrised}}. \label{eq:calf}
\end{align}
The second term in Equation \eqref{eq:fasq} is the so-called ``Lepage'' term.
The difference operator
acts on fermion fields as
\begin{equation}\label{eq:diffopquarks}
\nabla_{\mu}\psi(x) = \frac{1}{2a}\left[U_{\mu}(x)\psi(x+\hat{\mu}) -
U_{\mu}^{\dagger}(x-\hat{\mu})\psi(x-\hat{\mu})\right],
\end{equation}
whilst the discretised derivatives acting on link variables are, for
$\mu\neq\nu$,
\begin{align}
\nabla_{\mu}U_{\nu}(x) = {} &
\frac{1}{2}\Big[U_{\mu}(x)U_{\nu}(x+\hat{\mu})U_{\mu}^{\dagger}(x+\hat{\nu}
) \nonumber \\
{} & -
U_{\mu}^{\dagger}(x-\hat{\mu})U_{\nu}(x-\hat{\mu})U_{\mu}(x-\hat{\mu}
+\hat{\nu}) \Big],\label{eq:nablaU} \\
\nabla_{\mu}^{(2)}U_{\nu}(x) = {} &
\Big[U_{\mu}(x)U_{\nu}(x+\hat{\mu})U_{\mu}^{\dagger}(x+\hat{\nu}
) - 2U_{\nu}(x) \nonumber \\
{} & +
U_{\mu}^{\dagger}(x-\hat{\mu})U_{\nu}(x-\hat{\mu})U_{\mu}(x-\hat{\mu}
+\hat{\nu}) \Big].
\end{align}

The HISQ action is an extension of the AsqTad action that includes two levels of link fattening and
a tuned coefficient for the Naik term. Whilst the AsqTad action has
negligible tree level errors
for light
quarks, this is not true for charm or bottom quarks \cite{follana07}. Charm
quarks are
generally nonrelativistic in typical mesons, so the rest energy of the
quark is much larger than its momentum. The dominant tree level errors are therefore
${\cal O}(a^4m_0^4)$. One suppresses these errors by tuning the coefficient of the Naik term
\begin{equation}
\frac{a^2}{6}(\nabla_{\mu})^3\rightarrow
\frac{a^2}{6}(1+\epsilon)(\nabla_{\mu})^3.
\end{equation}

One also adds a second level of fattening in the link variables to reduce
the
discretisation errors arising from taste exchange interactions in the HISQ
action. Between the smearing operations, one sandwiches a reunitarisation
operator, ${\cal U}$, that projects the smeared link variables back to
$SU(3)$ or $U(3)$. For simplicity, the Lepage term is
included in the HISQ action only after the second level of link fattening. The resulting action is
\begin{equation}
S_{\text{HISQ}} =
a^4\sum_x\overline{\psi}(x)\left(\gamma_{\mu}\nabla_{\mu}^{
\text{HISQ}} + m_0\right)\psi(x),
\end{equation}
where
\begin{equation}
\nabla_{\mu}^{\text{HISQ}} = \nabla_{\mu}^{(FUF)} -
\frac{a^2}{6}(1+\epsilon)\left(\nabla_{\mu}^{(UF)}\right)^3.
\end{equation}
The superscripts indicate that the first operator,
$\nabla_{\mu}^{(FUF)}$, is built from the full HISQ-smeared links, given by
\begin{equation}
{\cal F}_{\mu}^{\text{HISQ}} = \left({\cal
F}_{\mu}- \sum_{\rho\neq\mu}\frac{a^2(\nabla_{\rho})^2}{2}\right){\cal U}{\cal
F}_{\mu},
\end{equation}
whilst the second operator, $\nabla_{\mu}^{(UF)}$, uses only one level of
smearing:
\begin{equation}
\nabla_{\mu}^{(UF)} = {\cal U}{\cal F}_{\mu}.
\end{equation}
We define the operator ${\cal F}_{\mu}$ in Equation \eqref{eq:calf}.

We give results for both massless and massive HISQ quarks. For massless
quarks the tuning parameter is just $\epsilon = 0$. For massive
quarks we set the tuning parameter to its tree level value, $\epsilon =
\epstree$, for consistency with non-perturbative simulations \cite{na12}.
We discuss this in more detail in Section \ref{sec:zqresults}.

\subsection{Heavy Quark Action}

For the heavy quark fields, $\psi(\mathbf{x},t)$, we use the NRQCD action of
\cite{gray05,dowdall12a}, which is improved through ${\cal O}(1/M_0^2)$
and ${\cal O}(a^2)$ and includes the leading relativistic ${\cal O}(1/M_0^3)$
correction. The NRQCD action is
\begin{align}
S_{\text{NRQCD}} = {} & 
\sum_{\mathbf{x},t}\psi^{\dagger}_{t}\psi_{t}
-\psi^{\dagger}_t\left(1-\frac{a\delta H}{2}\right)\left(1-
\frac{aH_0}{2n}\right)^n\nonumber \\
{} & \times U_4^{\dagger}\left(1-
\frac{aH_0}{2n}\right)^n \left(1-\frac{a\delta
H}{2}\right)\psi_{t-1},
\end{align}
where $\psi^{\dagger}_{t} = \psi^{\dagger}(\mathbf{x},t)$ and
$\psi_{t-1} = \psi(\mathbf{x},t-1)$. 

Here the leading kinetic term in the NRQCD action is given by
\begin{equation}
aH_0 = -\frac{\Delta^{(2)}}{2aM_0},
\end{equation}
and the correction terms are
\begin{align}
a\delta H = {} &
-c_1\frac{(\Delta^{(2)})^2}{8(aM_0)^3}+c_2\frac{i}{8(aM_0)^2}
\left(\nabla\cdot \widetilde{\mathbf{E}} - \widetilde{\mathbf{E}}\cdot
\nabla\right) \nonumber \\
{} & - c_3\frac{1}{8(aM_0)^2} \sigma\cdot
\left(\widetilde{\nabla}\times \widetilde{\mathbf{E}} -
\widetilde{\mathbf{E}}\times \widetilde{\nabla}\right) \nonumber \\
{} & - c_4\frac{1}{2aM_0}\sigma\cdot
\widetilde{\mathbf{B}}+c_5\frac{\Delta^{(4)}}{24aM_0}
-c_6\frac{(\Delta^{(2)})^2}{16n(aM_0)^2}.
\end{align}
All the derivatives are tadpole improved and the discretised difference
operators are
\begin{equation}
\Delta^{(2)} = \sum_{j=1}^2\nabla_j^{(2)}, \quad \Delta^{(4)} =
\sum_{j=1}^3\nabla_j^{(4)}, \quad \widetilde{\nabla}_i = \nabla_i - 
\frac{1}{6}\nabla_i^{(3)},
\end{equation}
where the improved operators act on fermion fields via
\begin{align}
\nabla_\mu^{(2)}\psi(x) = {} & U_\mu(x)\psi(x+\hat{\mu})\nonumber \\
{} & +  U^\dagger_\mu(x-\hat{\mu})\psi(x-\hat{\mu}) - 2\psi(x), \\
\nabla_\mu^{(3)}\psi(x) = {} & \frac{1}{2}\bigg[ U_\mu(x)U_\mu(x+\hat{\mu})\psi(x+2\hat{\mu})
\nonumber \\
{} & -  U^\dagger_\mu(x-\hat{\mu})U^\dagger_\mu(x-2\hat{\mu})\psi(x-2\hat{\mu})\bigg] \nonumber \\
{} & - U_\mu(x)\psi(x+\hat{\mu}) + U^\dagger_\mu(x-\hat{\mu})\psi(x-\hat{\mu}), \\
\nabla_\mu^{(4)}\psi(x) = {} & U_\mu(x)U_\mu(x+\hat{\mu})\psi(x+2\hat{\mu}),
\nonumber \\
{} & + U^\dagger_\mu(x-\hat{\mu})U^\dagger_\mu(x-2\hat{\mu})\psi(x-2\hat{\mu})+ 6\psi(x) \nonumber
\\
{} & - 4\Big(U_\mu(x)\psi(x+\hat{\mu}) + U^\dagger_\mu(x-\hat{\mu})\psi(x-\hat{\mu})\Big).
\end{align}
The improved chromo-electric and -magnetic fields,
$\widetilde{E}_j = \widetilde{F}_{i4}$ and $\widetilde{B}_j = -\epsilon_{ijk}\widetilde{F}_{jk}/2$,
are defined in terms of the improved field strength tensor, given by \cite{wingate03}:
\begin{align}
\widetilde{F}_{\mu\nu} = {} & \frac{5}{3}F_{\mu\nu}(x) -
\frac{1}{6}\bigg(U_\mu(x)F_{\mu\nu}(x+\hat{\mu})U^\dagger_\mu(x) \nonumber \\
{} & + U^\dagger_\mu(x-\hat{\mu})F_{\mu\nu}(x-\hat{\mu})U_\mu(x-\hat{\mu}) - (\mu\leftrightarrow
\nu)\bigg),
\end{align}
where
\begin{align}
F_{\mu\nu}(x) = {} & -\frac{i}{2g}\left(\Omega_{\mu\nu}(x) - \Omega_{\mu\nu}^\dagger(x)\right), \\
\Omega_{\mu\nu}(x) = {} & \frac{1}{4}\sum_{\{(\alpha,\beta)\}}U_\alpha(x)U_\beta(x+\hat{\alpha})
\nonumber \\
{} & \times U_{-\alpha}(x+\hat{\alpha}+\hat{\beta})U_{-\beta}(x+\hat{\beta}).
\end{align}
The final sum runs over 
\begin{equation}
\{(\alpha,\beta)\} = \{(\mu,\nu),(\nu,-\mu),(-\mu,-\nu),(-\nu,\mu)\},
\end{equation}
with $\mu\neq\nu$.

The values of the coefficients, $c_i$, in the NRQCD action are fixed
by matching lattice NRQCD to full QCD. We use the tree level values of
$c_i=1$ for all $i=1,\,\cdots,6$, and do not consider the effects of
radiative improvement of the NRQCD action.

\section{Quark self energy}

Perturbative calculations of the self energy for massless AsqTad quarks
were
carried out in \cite{gulez04} as part of the matching
calculation for NRQCD-AsqTad currents. In
this work, we extend these results to HISQ fermions. We update the results
for the massless case and generalise the results to massive quarks,
applying the methods of \cite{groote00} to extract the self energy
parameters.

\subsection{HISQ Parameters \label{ssec:hisqparms}}

The general formalism for self energy calculations is laid out in
\cite{mertens98} and developed in \cite{groote00}. In this section we
apply this formalism to the HISQ action, concentrating on the massive
case.

We start with the quark two-point correlation function,
\begin{equation}
\langle\psi(t,\mathbf{p}^{\prime})\overline{\psi}(0,\mathbf{p})\rangle =
(2\pi)^3\delta(\mathbf{p}-\mathbf{p}^{\prime})G(t,\mathbf{p}),
\end{equation}
which defines the quark propagator $G(t,\mathbf{p})$. The bare quark field
$\overline{\psi}(0,\mathbf{p})$ creates multiparticle states in addition
to a one-quark state and so one expects the quark propagator to take the
form
\begin{equation}
G(t,\mathbf{p}) = {\cal
Z}_2(\mathbf{p})e^{-E(\mathbf{p})t}\Gamma_{\text{proj}}+\cdots.
\end{equation}
Here $\Gamma_{\text{proj}}$ is a projection operator in Dirac space;
the ellipses represent multiparticle states and lattice artifacts, which
we will not consider any further; and ${\cal
Z}_2(\mathbf{p})$ is the single quark residue.

The use of a lattice regulator distorts the mass shell of the quark, which
would otherwise satisfy the relativistic dispersion relation in
Euclidean space. To account for the distorted pole position in a
systematic manner, one therefore defines the rest mass of the quark, $m_Q$,
as
\begin{equation}
m_Q = E(\mathbf{p}=\mathbf{0})
\end{equation}
and the wavefunction renormalisation as
\begin{equation}
Z_Q = {\cal Z}_2(\mathbf{p}=\mathbf{0}).
\end{equation}
In Sections \ref{sec:masslessres} and \ref{sec:massiveres} we will use
$Z_q$ and $Z_Q$ to denote the massless and massive wavefunction
renormalisations respectively; in this section, however, we use $Z_Q$
as shorthand for either $Z_q$ or $Z_Q$ for notational simplicity.

We renormalise at the point $(p_0,\mathbf{p}) = (iE,\mathbf{0})$
and therefore consider a zero spatial momentum quark propagating forward
in time, for which one expects
\begin{equation}\label{eq:propp0}
G(t,\mathbf{0}) = Z_Qe^{-Et}\frac{1+\gamma_0}{2}+\cdots.
\end{equation}

We denote the momentum space quark propagators for the full and free
theories $G(p)$ and $G_0(p)$ respectively. These propagators are
related via the quark self energy, $\Sigma(p)$:
\begin{equation}
G^{-1}(p) = G_0^{-1}(p) - \Sigma(p),
\end{equation}
where the self energy is the sum of all one-particle irreducible graphs;
in perturbation theory one assumes that the self energy is a ``small''
correction. The pole corresponding to the single particle quark state has
a nonzero residue in the limit that the self energy vanishes, whilst the
residues of the multiparticle states vanish in the absence of an
interaction.

Carrying out the Fourier transform in $p_0$ of the full quark propagator,
$G(p)$, one finds
\begin{equation}\label{eq:propt}
G(t,\mathbf{p}) = \int_{-\pi/a}^{\pi/a}
\frac{\text{d}p_0}{2\pi}e^{-ip_0t}G(p_0,\mathbf{p}).
\end{equation}
We identify this expression at zero spatial momentum with Equation
\eqref{eq:propp0}, which enables us to relate the mass and wavefunction
renormalisation to parameters in the action, via the quark propagator. In the
following
derivations, we will neglect factors of the lattice spacing $a$ for simplicity.
These can be easily
included at the end of the derivations by dimensional analysis.

\subsection{Pole mass}

For HISQ fermions, the form of the free propagator is
\begin{equation}
G_0^{-1}(p) =
\sum_{\mu}i\gamma_{\mu}\sin\left(p_{\mu}\right)K_{\mu}(p)+m_0.
\end{equation}
Here $m_0$ is the bare quark mass and
\begin{equation}
K_{\mu}(p) = 1+\frac{1+\epsilon}{6} \left(\sin p_{\mu}\right)^2.
\end{equation}
We write the one loop self energy as
\begin{equation}
\Sigma^{(1)}(p) =
\sum_{\mu}i\gamma_{\mu}\sin\left(p_{\mu}\right)\sigmaA_{\mu}
(p)
+ \sigmaB(p)\mathbb{I},
\end{equation}
where $\mathbb{I}$ is the identity element of the Clifford
algebra, so that the one loop propagator is
\begin{widetext}
\begin{equation}\label{eq:propfull}
G(p) =
\frac{-\sum_{\mu}i\gamma_{\mu}\sin\left(p_{\mu}\right)\left[K_{\mu}(p)
- \alpha_s\sigmaA_{\mu}(p)\right] + m_0 -
\alpha_s\sigmaB(p)}{\sum_{\rho}\left(\sin\left(p_{\rho}
\right)\right)^2\left[K_{\rho}(p) -
\alpha_s\sigmaA_{\rho}(p)\right]^2+\left[m_0 -
\alpha_s\sigmaB(p)\right]^2}.
\end{equation}
\end{widetext}

At zero spatial momentum the pole condition for
the forward propagating quark is
\begin{equation}\label{eq:polecondition}
\sinh(E)\left(1-\frac{1+\epsilon}{6}\left(\sinh(E)\right)^2 -
\alpha_s\sigmaA_0\right) = m_0 - \alpha_s\sigmaB,
\end{equation}
where we have neglected the arguments of $\sigmaA$ and $\sigmaB$ for
clarity. We now expand the quark energy and tuning parameter
$\epsilon$ to
one loop
as
\begin{align}
E = {} &  \mtree + \als m_1, \label{eq:mexpansion}\\
\epsilon = {} & \epstree + \als\epsilon_1. \label{eq:epsilonexpansion}
\end{align}

Substituting these expressions into the pole condition, Equation
\eqref{eq:polecondition}, gives an expression for the tree level
pole mass, $\mtree$, at fixed bare mass, $m_0$:
\begin{equation}\label{eq:mtreedef}
\sinh(\mtree)\left[1-\frac{1+\epstree}{6}\left(\sinh(\mtree)\right)^2
\right] = m_0.
\end{equation}
We then fix $\epstree$ by requiring that the tree level pole
mass is equal to the tree level kinetic mass. We discuss this condition in
more detail in
Appendix \ref{app:oneloop}. One ultimately finds
\begin{align}\label{eq:epstreedef}
\epstree = {} & -1 +  
\frac{1}{\left(\sinh(\mtree)\right)^2}
\nonumber \\
{} & \times \left[4-\sqrt{4+\frac{12\mtree}{
\cosh(\mtree)
\sinh(\mtree)}}\right].
\end{align}
Expanding this equation gives Equation (24) of
\cite{follana07}. We obtain
a precise numerical value for the tree level mass by solving Equations
\eqref{eq:mtreedef} and \eqref{eq:epstreedef} self-consistently; we find that
a series solution is insufficiently accurate for our accurately setting the
light quarks onshell. 

We repeat the process at one loop to obtain
\begin{equation}\label{eq:m1}
m_1 = Z_Q^{(0)}\left\{\frac{\epsilon_1}{6}\left(\sinh(\mtree)\right)^3
+ \sinh(\mtree)\sigmaA_0 -
\sigmaB\right\},
\end{equation}
where $Z_Q^{(0)}$ is the tree level wavefunction renormalisation, given by
\begin{equation}\label{eq:zqtree}
Z_Q^{(0)} =
\left\{\cosh(\mtree)\left[1-\frac{1+\epstree}{2}
\left(\sinh(\mtree)\right)^2
\right] \right\}^{-1}.
\end{equation}

\subsection{Wavefunction renormalisation}

We now extract the wavefunction renormalisation from the quark propagator.
Recall that the wavefunction renormalisation is given by residue of the single
particle momentum pole obtained by identifying Equations \eqref{eq:propp0} and
\eqref{eq:propt} at zero spatial momentum, whence
\begin{equation}\label{eq:gdefforz}
\int_{-\pi/a}^{\pi/a}
\frac{\text{d}p_0}{2\pi}e^{-ip_0t}G(p_0,\mathbf{0}) =
Z_Qe^{-Et}\frac{1+\gamma_0}{2}+\cdots.
\end{equation}

It is convenient to re-express this relation in terms of the variable $z =
e^{ip_0}$:
\begin{equation}
\int_{-\pi/a}^{\pi/a} \frac{\text{d}p_0}{2\pi}e^{-ip_0t}G(p_0,\mathbf{0})
\stackrel{p_0\rightarrow z}{=}-i\oint
\frac{\text{d}z}{2\pi}z^{t-1}G(z,\mathbf{0}),
\end{equation}
where the contour of integration is now around the unit circle in the
complex $z$-plane. Writing the propagator as $G(z,\mathbf{0}) =
g_1(z)/g_2(z)$, the residue at $z = z_1 = e^{-E}$ is
\begin{equation}
\res_{\;z=z_1}\left\{z^{t-1} G(z,\mathbf{0})\right\} =
z_1^t\frac{g_1(z_1)}{z_1g_2^{\prime}(z_1)},
\end{equation}
where the prime indicates differentiation with respect to $z$.

In this case the quark propagator is given by Equation \eqref{eq:propfull}
and we obtain
\begin{widetext}
\begin{equation}
\res_{\;z=z_1}\left\{z^{t-1} G(z,\mathbf{0})\right\} = e^{-Et}
\frac{1+\gamma_0}{2}\Bigg\{\cosh(E)\left[1 -
\frac{1+\epsilon}{2}(\sinh(E))^2\right]+\alpha_si\frac{\text{d}}{\text{d}p_0}
\Big[
i\sin(p_0)\sigmaA_0 + \sigmaB\Big]\Bigg\}^{-1}.
\end{equation}
Comparing this equation with Equation \eqref{eq:gdefforz}, we read off the
wavefunction renormalisation as
\begin{equation}
Z_Q = \Bigg\{\cosh(E)\left[1 -
\frac{1+\epsilon}{2}(\sinh(E))^2\right]+
\alpha_si\frac{\text{d}}{\text{d}p_0}\Big[i\sin(p_0)\sigmaA_0 +
\sigmaB\Big]\Bigg\}^{-1}.
\end{equation}

We again expand the mass and tuning parameter as in Equations
\eqref{eq:mexpansion} and \eqref{eq:epsilonexpansion}. The tree level
result reduces to Equation \eqref{eq:zqtree}, whilst the one loop
wavefunction renormalisation is
\begin{align}\label{eq:z1def}
Z_Q^{(1)} = {} & Z_Q^{(0)}\bigg\{\frac{\epsilon_1}{2}
\cosh(\mtree)\left(\sinh(\mtree)\right)^2
-m_1\sinh(\mtree)\bigg[1-\frac{1+\epstree}{2}
\Big(2\left(\cosh(\mtree)\right)^2
+\left(\sinh(\mtree)\right)^2\Big) \bigg] \nonumber \\
{} & +\frac{\text{d}}{\text{d}p_0}
\Big[\sin(p_0)\sigmaA_0-i\sigmaB\Big]\bigg\}.
\end{align}
Here we have found it convenient to factor out the tree level wavefunction
renormalisation to ensure the one loop terms have the correct infrared
divergence \cite{groote00,mertens98}. In other words, we set
\begin{equation}
\frac{Z_Q}{Z_Q^{(0)}} = 1+\alpha_sZ_Q^{(1)}+{\cal O}(\alpha_s^2).
\end{equation}
\end{widetext}

\subsection{\label{sec:zqresults}Numerical Results}

\begin{figure}
\includegraphics[height=0.08\textwidth,width=0.3\textwidth]{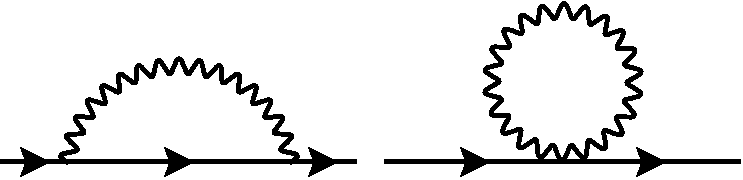}
\caption{\label{fig:zqfeyndiags}Contributions to the one loop self energy
required for the HISQ quark mass and wavefunction renormalisation. On the
left is the ``rainbow'' diagram and on the right the ``tadpole'' diagram. 
Straight lines represent light quarks and curly lines indicate gluons.}
\end{figure}

In this section we summarise our results for both massless and
massive HISQ quarks. The diagrams that contribute to the self energy at
one loop are shown in Figure \ref{fig:zqfeyndiags}. The continuum-like
contribution is the ``rainbow'' diagram, shown on the left of Figure
\ref{fig:zqfeyndiags}. On the right is the lattice artifact ``tadpole''
diagram. We calculated the corresponding Feynman integrals
using two independent methods.

Our first method employed the automated lattice perturbation theory
routines \hippy and \hpsrc \cite{hart04,hart09}. These routines have
now been used in a number of perturbative calculations, for example in
\cite{drummond02,drummond03a,drummond03b,hart04,hart07,mueller11,
hammant11,dowdall12a,dowdall12b}, and have been extensively tested against
results
published
in the literature.

Evaluating the relevant Feynman integrals is a two stage process: we
first generate the Feynman rules with \hippycomma a set of \python routines
that encode the Feynman rules in ``vertex files''. These vertex files are
then read in by the
\hpsrc code, which is a collection of \fortran modules that reconstruct
the diagrams and evaluates the corresponding integrals numerically, using
the \vegas algorithm \cite{lepage80}. All derivatives
of the self energy are implemented analytically using 
the derived \verb+taylor+ type, defined as part of the \fortran
\verb+TaylUR+ package \cite{hippel10}. We performed our calculations on
the Darwin cluster at the Cambridge High Performance
Computing Service, as part of the DiRAC facility, and the
Sporades cluster at the College of William
and Mary with routines adapted for
parallel computers using MPI (Message Passing Interface).

In contrast
to previous matching calculations, such as \cite{gulez04}, we do not
attempt to present
Feynman rules for the improved NRQCD and massive HISQ actions: the
automated lattice
perturbation theory procedure does not require such explicit analytic
expressions. This
method therefore reduces the possibility of algebraic errors in the
manipulation of
Feynman integrands.

We undertook a number of tests of our automated perturbation theory code. In
particular, we reproduced
the results of \cite{gulez04} with massless AsqTad light quarks and NRQCD 
heavy quarks. The chief advantage of the automated lattice perturbation
theory routines is the
relative ease with which different actions can be implemented in the
calculation. Once the correct \hpsrc code is in place to calculate the
requisite Feynman
diagrams, switching actions is just a matter of replacing the input
vertex files generated
by \hippydot

In many cases, we established that gauge invariant
quantities, such as the mass renormalisation, are gauge parameter
independent by working in both Feynman and Landau gauges.

Furthermore we confirmed that infrared
divergent quantities, such as the wavefunction renormalisation, exhibited
the correct continuum-like behaviour. We regulate the infrared behaviour
with a gluon mass for 24 different values of the gluon mass between
$a^2\lambda^2 = 10^{-7}$ and $a^2\lambda^2 = 10^{-12}$. Fitting these
results to a logarithmic function establishes that the code
correctly reproduced the expected logarithmic behaviour. To extract the
infrared finite piece of infrared
divergent quantities we constrain the fit function to have the
correct logarithmic divergence.

At finite lattice spacing offshell contributions to the vertex
renormalisation must be removed to restore the correct continuuum-like
infrared behaviour. We set the HISQ quarks exactly onshell and remove offshell
contributions to the vertex renormalisation with an onshell
projector.
This corresponds to imposing the equation of motion on the quark or
antiquark spinor, just as would be done analytically \cite{kuramashi98}. It
is important to ensure the quark is set exactly onshell, by solving the
full inverse tree level HISQ propagator for the timelike component of the
quark momentum, or the continuum infrared behaviour is not recovered. 
We found that this requires
very precise values for $\mtree$ and $\epstree$ 
(see Table \ref{tab:hisqmparms}) and that using only a few digits is
insufficient.
Likewise the equation of motion for the massive HISQ propagator must be
exactly satisfied for the offshell contributions to be fully removed.

Our second method is based on \mathematica and \fortran 
routines developed previously for matching of NRQCD/AsqTad currents 
\cite{gulez04} and adapted here for HISQ light quarks.  Although Feynman rules
for 
one- and two-gluon emission vertices are known for the 
NRQCD and AsqTad actions and are used in the present calculations as well, 
the HISQ vertices needed to be handled differently.  Analytic expressions 
for HISQ vertices are too complicated to write down in closed form. 
Instead we build up one- and two-gluon emission vertices emerging from the 
HISQ action from vertices of simpler operators through repeated use 
of convolution rules \cite{morningstar93}.  
For instance, since one- and two-gluon emission vertices are known for once 
fattened links from the AsqTad Feynman rules, 
one can use them to build up vertices of a product of 
three, five, seven such fat links and implement the second fattening.  

We use \mathematica to carry out all the Dirac algebra and also to 
take derivatives of NRQCD vertices with respect to external momenta.  
We have developed \fortran ``automatic differentiation'' routines to take 
derivatives of HISQ vertices.  The same bookkeeping 
used for repeated application of convolution rules allows us here to 
apply the chain rule of differentiation each time 
two expressions are multiplied and build up derivatives of the 
complicated HISQ vertices.

In our second method the correct infrared singularities were isolated 
and in many cases handled with subtraction functions.  Details of the
subtraction 
functions are given in Appendix \ref{app:subfn}.

We believe that these two methods are sufficiently independent that, in
conjunction with tests of gauge invariance and correct infrared behaviour
and the replication of results in the literature,
agreement between these methods provides a stringent check of our results.

We now give our numerical results for the HISQ quark mass and wavefunction
renormalisation.

\subsubsection{\label{sec:masslessres}Massless Quarks}

For massless quarks we require only the wavefunction renormalisation. In
this case $a\mtree=\epstree=0$ and so Equation \eqref{eq:z1def} reduces to
\begin{equation}
Z_q^{(1)} = - i\frac{\text{d}}{\text{d}(ap_0)}
\Big[i\sin(ap_0)\sigmaA_0+\sigmaB\Big].
\end{equation}

The wavefunction renormalisation is infrared divergent and we decompose
our results into an infrared finite contribution, $C_q$, and an infrared
divergent contribution, $C_q^{\text{IR}}$. Thus we write
\begin{align}
Z_q = {} & 1+\als Z_q^{(1)} +{\cal O}(\als^2) \nonumber \\
= {} & 
1+\als\left(C_q^{\text{IR}}+C_q\right)+{\cal O}(\als^2).
\end{align}
The infrared divergence is given by
\begin{equation}
C_q^{\text{IR}} = 
\frac{1}{3\pi}\left[1+\left(\xi-1\right)\right]
\log\left(a^2\lambda^2\right),
\end{equation}
where $\lambda$ is the gluon mass, introduced to regulate the infrared
behaviour, and $\xi$ is the gauge parameter. For massless quarks
the infrared divergences in the lattice
matching coefficients, arising from the wavefunction and vertex
renormalisations, are ultimately cancelled by corresponding
divergences in continuum QCD. We confirm that any gluon
mass dependence cancels between the lattice and continuum one loop
coefficients.

In contrast to the AsqTad and NRQCD actions, we do not need to use
tadpole improvement for HISQ and the only contributions to the infrared
finite piece, $C_q$, are the rainbow and tadpole diagrams,
\begin{equation}
C_q = C_q^{\text{rbow}} + C_q^{\text{tad}}.
\end{equation}

We tabulate our results for the wavefunction renormalisation in Table
\ref{tab:hisqm0parms}.

\begin{table}
\caption{\label{tab:hisqm0parms}One-loop wavefunction renormalisation for
massless HISQ fermions. The gauge parameter is $\xi$. All uncertainties are
statistical
errors arising from the numerical integration of the relevant diagrams. \\}
\begin{ruledtabular}
\begin{tabular}{cccc}
\vspace*{-5pt}\\
$\xi$ & $C_q^{\text{rbow}}$ & $C_q^{\text{tad}}$ & $C_q$ \\
\vspace*{-5pt}\\
\hline
\vspace*{-5pt}\\
1 & -0.8183(1) & 0.4243(3) & -0.3940(3) \\
0 & -0.0198(1) & 0.1343(3) & 0.1145(3) \\
\vspace*{-5pt}\\
\end{tabular}
\end{ruledtabular}
\end{table}

\subsubsection{\label{sec:massiveres}Massive Quarks}

We require both the mass and wavefunction
renormalisation for massive HISQ fermions. In general, both of these
quantities are
functions of $\epsilon_1$. For consistency with the HISQ action
used in numerical simulations, however, we ignore $\epsilon_1$
in Equations \eqref{eq:m1} and \eqref{eq:z1def}.
In Reference \cite{follana07} it was found that the nonperturbatively determined
values for $\epsilon$ were always close to $\epstree$.  This justified
ignoring one-loop (or higher order)
corrections to $\epstree$ in all subsequent numerical simulations with
massive HISQ quarks. Perturbative matching that is going to be combined with
numerical computations must be consistent with the latter. We
set $\epsilon = \epstree$ accordingly.

Neglecting $\epsilon_1$ considerably
simplifies the perturbative calculation of both $am_1$ and $Z_Q^{(1)}$. For
completeness we tabulate our results
for $\epstree$, $a\mtree$ and $am_1$ in Table
\ref{tab:hisqmparms}. We
present results for a range of quark masses corresponding to the MILC
ensembles used in \cite{na12}, \cite{dowdall12a} and \cite{dowdall12b}. 
\begin{table}
\caption{\label{tab:hisqmparms}Tree level and one loop tuning
parameters for massive HISQ fermions. All uncertainties are statistical
errors arising from the numerical integration of the relevant diagrams. \\}
\begin{ruledtabular}
\begin{tabular}{ccccc}
\vspace*{-5pt}\\
$am_0$ & $\epstree$ & $a\mtree$ & $am_1$ \\
\vspace*{-5pt}\\
\hline
\vspace*{-5pt}\\
0.826\hphantom{0} & -0.344960900736 & 0.814526131431 & 0.6580(1) \\
0.818\hphantom{0} & -0.340115648115 & 0.807017346575 & 0.6551(1) \\
0.645\hphantom{0} & -0.234829780198 & 0.641330413102 & 0.5871(1) \\
0.6300 & -0.225853340666 & 0.626715862647 & 0.5811(1) \\
0.627\hphantom{0} & -0.224064962178 & 0.623789107649 & 0.5795(1) \\
0.6235 & -0.221981631663 & 0.620372982565 & 0.5784(1) \\
0.6207 & -0.220317446966 & 0.617638873348 & 0.5771(1) \\
0.434\hphantom{0} & -0.117189612523 & 0.433453860575 & 0.4855(1) \\
0.4130 & -0.106941294689 & 0.412571424109 & 0.4734(1) \\
0.4120 & -0.106461983347 & 0.411576478677 & 0.4728(1) \\
\vspace*{-5pt}\\
\end{tabular}
\end{ruledtabular}
\end{table}

The one loop mass renormalisation is gauge invariant and infrared finite,
whilst the wavefunction renormalisation is gauge dependent and infrared
divergent. We write the one loop
wavefunction
renormalisation in Equation \eqref{eq:z1def} as
\begin{equation}
Z_Q^{(1)} =  Z_Q^{(m_1)}am_1 + Z_Q^{(\Sigma)},
\end{equation}
where
\begin{align}
& Z_Q^{(m_1)} = -Z_Q^{(0)}\sinh(a\mtree) \, \times\nonumber \\
& \bigg[1-\frac{(1+\epstree)}{2}
\Big(2\left(\cosh(a\mtree)\right)^2
+\left(\sinh(a\mtree)\right)^2\Big) \bigg], \label{eq:zqm1}\\
 & \;Z_Q^{(\Sigma)}= {}Z_Q^{(0)}\frac{\text{d}}{\text{d}(ap_0)}
\Big[\sin(ap_0)\sigmaA_0-i\sigmaB\Big].
\end{align}
Recall that we have set $\epsilon_1=0$. The contribution from
$Z_Q^{(\Sigma)}$ contains the logarithmic infrared
divergence. In line with our presentation of the massless case, we
separate the infrared finite and divergent pieces of the one loop self
energy-dependent contribution,
which
we denote $C_Q$ and $C_Q^{\text{IR}}$ respectively. Thus we have
\begin{equation}\label{eq:zQdef}
Z_Q^{(1)} = C_Q +C_Q^{\text{IR}},
\end{equation}
where the infrared divergent contribution is
given by
\begin{equation}
C_Q^{\text{IR}} = 
\frac{1}{3\pi}\left[-2+\left(\xi-1\right)\right]
\log\left(a^2\lambda^2\right).
\end{equation}
We further decompose the infrared finite contribution into the self energy
rainbow and
tadpole diagram and $m_1$-dependent pieces:
\begin{equation}
C_Q = Z_Q^{(m_1)}am_1 + C_Q^{\text{rbow}} + C_Q^{\text{tad}}.
\end{equation}

We give our results for the one loop wavefunction renormalisation in
Feynman
gauge in Table \ref{tab:hisqmparmsZQfeyn}.
\begin{table}
\caption{\label{tab:hisqmparmsZQfeyn}One-loop wavefunction renormalisation
for massive HISQ fermions. All results in Feynman gauge. Quoted
uncertainties are statistical errors from the numerical integration
of the relevant diagrams. The quantity $Z_Q^{(m_1)}$, defined in Equations
\eqref{eq:zqm1} and \eqref{eq:zqtree}, is an analytic function of only
$\epstree$ and $a\mtree$. These parameters, given in Table
\ref{tab:hisqmparms}, are known to twelve significant figures; we therefore
neglect the uncertainty in $Z_Q^{(m_1)}$ here.\\}
\begin{ruledtabular}
\begin{tabular}{ccccc}
\vspace*{-5pt}\\
$am_0$ & $C_Q^{\text{rbow}}$ & $C_Q^{\text{tad}}$ & $Z_Q^{(m_1)}$ & $C_Q$ \\
\vspace*{-5pt}\\
\hline
\vspace*{-5pt}\\
0.826\hphantom{0} & -1.342(1) & 0.1952(1) & 0.427495 & -0.865(1) \\
0.818\hphantom{0} & -1.349(1) & 0.1989(1) & 0.415945 & -0.878(1) \\
0.645\hphantom{0} & -1.510(1) & 0.2888(1) & 0.210922 & -1.097(1) \\
0.6300 & -1.511(1) & 0.2949(1) & 0.197029 & -1.102(1) \\
0.627\hphantom{0} & -1.530(1) & 0.2970(1) & 0.194322 & -1.120(1) \\
0.6235 & -1.534(1) & 0.2981(1) & 0.191192 & -1.125(1) \\
0.6207 & -1.537(1) & 0.2982(1) & 0.188712 & -1.130(1) \\
0.434\hphantom{0} & -1.785(1) & 0.3652(1) & 0.066076 & -1.388(1) \\
0.4130 & -1.820(1) & 0.3715(1) & 0.057058 & -1.421(1) \\
0.4120 & -1.821(1) & 0.3712(1) & 0.056650 & -1.423(1) \\
\vspace*{-5pt}\\
\end{tabular}
\end{ruledtabular}
\end{table}

\subsection{\label{sec:nrqcdparms}NRQCD Parameters}

The one loop parameters of NRQCD have been extensively studied in the
literature, for example in \cite{morningstar93,gulez04,dowdall12a}. Indeed,
a two loop calculation of the energy shift has recently been carried out with
a mixed approach that combines automated lattice perturbation theory
calculations of the fermionic contributions with results extracted
from quenched weak coupling simulations for all other
contributions \cite{hart12}. Here we simply introduce the notation and summarise
the
necessary results at the heavy quark masses relevant for the simulations in
\cite{na12}. We require the wavefunction renormalisation, $Z_H$, and the
mass renormalisation, $Z_M$:
\begin{align}
Z_H = {} & 1 + \als\left(C_H^{\text{IR}}+C_H\right) + {\cal
O}(\als^2), \\
Z_M = {} & 1 + \als C_M + {\cal O}(\als^2).
\end{align}
The infrared behaviour of NRQCD must match that of full
QCD and is therefore just
\begin{equation}
C_H^{\text{IR}} = \frac{1}{3\pi}\left[-2+\left(\xi-1\right)\right]
\log\left(a^2\lambda^2\right).
\end{equation}

In this case the infrared
finite contribution, $C_H$, is composed solely of the heavy
quark rainbow diagram, because both the tadpole diagram and tadpole
improvement contribution vanish \cite{gulez04}. The  mass renormalisation,
on the other hand, depends on both the rainbow and tadpole diagrams and the
tadpole improvement term,
\begin{equation}
C_M = C_M^{\text{rbow}}+C_M^{\text{tad}}+C_M^{u_0},
\end{equation}
where an analytic expression for $C_M^{u_0}$ is given in \cite{gulez04}:
\begin{align}
C_M^{u_0} = {} & \bigg[-1+\frac{3}{2n(aM_0)}+\frac{c_5}{3} \nonumber \\
{} & -
3(aM_0)\left(\frac{c_1}{(aM_0)^3}+\frac{c_6}{2n(aM_0)^2}\right)\bigg]
u_0^{(1)} .
\end{align}
At one loop we need not distinguish
between the pole mass and the bare mass, so for convenience we express all
results in terms of the bare mass. 

We tabulate our results for $C_H$ and $C_M$ in Table
\ref{tab:nrqcdparms}. We present results with $c_i = 1$ and use the Landau
link definition of the tadpole improvement factor $u_0$, with
$u_0^{(1)}=0.7503(1)$. All results use
stability parameter $n=4$.

\begin{table}
\caption{\label{tab:nrqcdparms}One-loop heavy quark
parameters in NRQCD.
All results use stability parameter $n=4$. We implement tadpole improvement
with the Landau link definition of $u_0$. All results are in Feynman gauge.
The quoted uncertainties are statistical
errors from the numerical integration of the relevant diagrams. \\}
\begin{ruledtabular}
\begin{tabular}{ccc}
\vspace*{-5pt}\\
$aM_0$ & $C_H$ & $C_M$ \\
\vspace*{-5pt}\\
\hline
\vspace*{-5pt}\\
3.297 & -0.235(1) & 0.167(1) \\
3.263 & -0.241(1) & 0.176(1) \\
3.25\hphantom{0} & -0.244(1) & 0.178(1) \\
2.688 & -0.362(1) & 0.262(1) \\
2.66\hphantom{0} & -0.366(1) & 0.264(1) \\
2.650 & -0.371(1) & 0.267(1) \\
2.62\hphantom{0} & -0.374(1) & 0.272(1) \\
1.91\hphantom{0} & -0.617(1) & 0.434(1) \\
1.89\hphantom{0} & -0.627(1) & 0.448(1) \\
1.832 & -0.657(1) & 0.466(1) \\
1.826 & -0.660(1) & 0.468(1) \\
\vspace*{-5pt}\\
\end{tabular}
\end{ruledtabular}
\end{table}

\section{\label{sec:matching}The Matching Procedure}

In lattice QCD the axial-vector and vector
current operators mix with higher order operators under renormalisation.
In this section we outline the perturbative matching procedure that
relates the lattice and continuum currents and the
extraction of the one loop mixing matrix elements.

Our strategy for the perturbative matching of heavy-light currents with
massless relativistic quarks and NRQCD heavy quarks follows that developed
in
\cite{morningstar98,morningstar99} and outlined in \cite{gulez04}. We will
briefly review the matching formalism and refer the reader to the earlier
articles. A related matching calculation for massless HISQ quarks with
NRQCD formulated in a moving frame (mNRQCD) was undertaken for the vector
and tensor heavy-light currents in \cite{mueller11}.

For massive quarks similar matching calculations using the same
lattice action for both quarks have been carried
out for Wilson quarks in \cite{kuramashi98} and for various
implementations of NRQCD in
\cite{braaten95,jones99,boyle00,hart07}. To our
knowledge, no matching
calculations with mixed actions and massive relativistic quarks have
been reported in the literature. 

Moving from massless to massive relativistic quarks complicates the
matching procedure. In the former case, quarks and antiquarks at zero
spatial momentum are indistinguishable and consequently scattering and
annihilation processes give
identical results. In the massive case, however, we must distinguish between
quarks and antiquarks. For HISQ quarks at zero spatial momentum, this
corresponds to
choosing $ap_0 = ia\mtree$ or $ap_0=-ia\mtree$ respectively. We choose
outgoing quarks or antiquarks --- the ``scattering'' or ``annihilation''
channels respectively --- to ensure we do not attempt to compute vanishing
matrix elements. Thus we calculate the matrix elements of $V_0$ and $A_k$
in the scattering channel and of $A_0$ and $V_k$ in the annihilation
channel. This procedure is valid, even at nonzero lattice spacing,
provided we match to the same channel in continuum QCD.

Unfortunately, from the practical viewpoint of calculating Feynman
diagrams, using massive quarks complicates the numerical
integration considerably. The chief difficulty lies in the annihilation
channel, which contains a
Coulomb singularity that must be handled with a subtraction
function. We discuss the subtraction functions employed in this work in
more detail in Appendix \ref{app:subfn}. Furthermore, in the automated
perturbation theory
routines, the pole in the NRQCD propagator crosses the contour of integration
and we can no longer carry out the usual Wick rotation back to Minkowski
space. Instead, we must deform the integration contours and
introduce a triple contour to ensure the stability of
numerical integration
\cite{hart07,mueller11}.

For both channels, the lattice matrix elements must be matched to their
continuum QCD counterparts. Analytic expressions for the relevant QCD
contributions already exist in the literature. References \cite{braaten95}
and \cite{jones99} discuss the annihilation channel for the axial-vector
current, whilst \cite{boyle00} present results for both vector and
axial-vector currents in the scattering channel at nonzero spatial
momentum. Results for components of both currents at zero spatial momentum
in both channels are presented in \cite{kuramashi98}. Whilst the authors of
\cite{hart07} are also concerned with calculating
matching coefficients for the spacelike components of the vector
current for lattice NRQCD, a procedure conceptually similar to that
discussed in this work, they take a slightly different approach,
calculating the continuum integrals numerically.

We calculate the mixing matrix
elements required
to match the axial-vector and vector currents in the effective NRQCD
theory to full QCD for the following combinations of
currents, Lorentz indices and orders in the  perturbative and $1/M_0$ expansions:
\begin{enumerate}
\item massless relativistic quarks:
\begin{enumerate}
\item $V_0$ through\\ ${\cal O}(\als,\Lambda_{\text{QCD}}/M_0,\als/(aM_0),\als
\Lambda_{\text{QCD}}/M_0)$;
\item $V_k$ ($k=1,2,3$) through\\ ${\cal O}(\als,\Lambda_{\text{QCD}}/M_0,\als/(aM_0))$;
\end{enumerate}
\item massive relativistic quarks:
\begin{enumerate}
\item $V_\mu$ ($\mu=1,\ldots,4$) through\\ ${\cal O}(\als,\Lambda_{\text{QCD}}/M_0,\als/(aM_0))$;
\item $A_\mu$ through ${\cal O}(\als,\Lambda_{\text{QCD}}/M_0,\als/(aM_0))$.
\end{enumerate}
\end{enumerate}
The results for both axial-vector and vector currents are
identical for massless relativistic quarks. To simplify our presentation we
therefore only give results for the vector current for massless
HISQ quarks.

We discuss each of these cases in turn.

\subsection{Massless Quarks}

\subsubsection{Temporal vector current}

We require three currents to match the temporal component of the vector current on the
lattice to full QCD through ${\cal
O}(\als,\Lambda_{\text{QCD}}/M_0,\als/(aM_0),\als\Lambda_{\text{QCD}}/M_0)$. These are
\begin{align}
J_{\mu}^{(0)}(x) = {} & \overline{q}(x)\Gamma_{\mu}Q(x), \label{eq:j0def}\\
J_{\mu}^{(1)}(x) = {} &
-\frac{1}{2(aM_0)}\overline{q}(x)\Gamma_{\mu}\mathbf{\gamma}
\cdot \overrightarrow{\mathbf{\nabla}} Q(x), \label{eq:j1def} \\
J_{\mu}^{(2)}(x) = {} & 
-\frac{1}{2(aM_0)}\overline{q}(x)\mathbf{\gamma}\cdot 
\overleftarrow{\mathbf{\nabla}}
\gamma_0 \Gamma_{\mu}Q(x). \label{eq:j2def}
\end{align}
Here the $Q$ fields are four component Dirac spinors with the upper two
components given by the two component NRQCD field and lower components
equal to zero. The $\Gamma_{\mu}$ operator represents the
vector current operator, so that here we have $\Gamma_{\mu} =
\gamma_{\mu}$. The difference operator $\nabla$ is defined
in Equation \eqref{eq:diffopquarks}, with the arrow indicating whether the
operator acts to the left or right. The Euclidean gamma matrices obey
\begin{equation}
\{\gamma_{\mu},\gamma_{\nu}\} = 2\delta_{\mu\nu},\qquad \gamma_\mu^\dagger
= \gamma_\mu.
\end{equation}

The matrix element of the timelike vector current in full QCD is related to the matrix elements
of the currents in the effective theory via
\begin{align}
\langle V_0 \rangle = {} & \left(1+\als \rho_0^{\,(V_0)}\right)
\langle J_0^{(0)}\rangle 
+ \left(1+\als \rho_1^{\,(V_0)}\right)\langle
J_0^{(1),\,\text{sub}}\rangle \nonumber \\
{} & + \als \rho_2^{\,(V_0)}\langle J_0^{(2),\,\text{sub}}\rangle + {\cal
O}(\als^2,\Lambda_{\text{QCD}}^2/M_0^2,a^2\als).\label{eq:jv0}
\end{align}
Here we have expressed the lattice currents in terms of the subtracted
currents,
\begin{equation}
J_\mu^{(i),\,\text{sub}} = J_\mu^{(i)} - \als \zeta_{10}J_\mu^{(0)}
\end{equation}
for $i = 1,\,2$. The subtracted currents are more physical and have
improved
power law behaviour \cite{collins01}.

The matching coefficients are given by
\begin{align}
\rho_0^{\,(V_0)} = {} & B_0^{\,(V_0)} - \frac{1}{2}(C_q+C_H) -
\zeta_{00}^{\,(V_0)}, \label{eq:rho0V0} \\
\rho_1^{\,(V_0)} = {} & B_1^{\,(V_0)} - \frac{1}{2}(C_q+C_H) - C_M -
\zeta_{01}^{\,(V_0)} - \zeta_{11}^{\,(V_0)}, \label{eq:rho1V0} \\
\rho_2^{\,(V_0)} = {} & B_2^{\,(V_0)} - \zeta_{02}^{\,(V_0)} -
\zeta_{12}^{\,(V_0)},\label{eq:rho2V0} 
\end{align}
where the $B_i$ arise from the matrix elements in full QCD and are given by
\cite{morningstar98,morningstar99,gulez04}
\begin{align}
B_0^{\,(V_0)} = {} & \frac{1}{\pi}\left(\ln(aM_0) - \frac{1}{4}\right), \\
B_1^{\,(V_0)} = {} & \frac{1}{\pi}\left(\ln(aM_0) - \frac{19}{12}\right),
\\
B_2^{\,(V_0)} = {} & \frac{4}{\pi}.
\end{align}
The renormalisation parameters $C_q$, $C_H$ and $C_M$ are the one loop
self energy corrections discussed in the previous sections. For convenience
we have
written the pole mass, which is common to both lattice and continuum
theories, in terms of the bare quark mass. We must
therefore include the one loop mass renormalisation that relates
these two masses in the tree level
contribution from $J_0^{(1)}$.

\begin{figure}
\includegraphics[height=0.20\textwidth,width=0.3\textwidth]{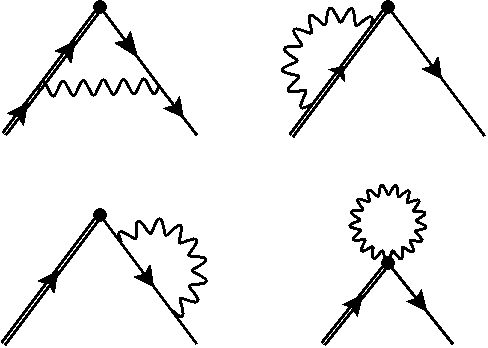}
\caption{\label{fig:zetaijfeyndiags}Contributions to the one loop
mixing matrix elements to match the vector and axial-vector currents in lattice NRQCD
to full QCD. Clockwise from top left to lower left are the
``vertex correction'' diagram, the ``heavy
earlobe'' diagram, ``vertex
tadpole'' diagram and the ``light earlobe'' diagram. The double lines
indicate
heavy quarks, the single lines represent light quarks and the wavy lines are gluons. Operator
insertions are denoted by the solid circles.}
\end{figure}

The $\zeta_{ij}^{\,(V_0)}$ in Equations \eqref{eq:rho0V0} to
\eqref{eq:rho2V0} are the one loop mixing matrix elements that arise from
the mixing of the currents. These contributions are generated by the one
loop diagrams in Figure \ref{fig:zetaijfeyndiags}. From top left to lower left
these are: the ``vertex correction'' diagram, the ``heavy earlobe''
diagram, the ``vertex tadpole''
diagram and the ``light earlobe'' diagram. To extract the mixing matrix
elements, we insert one of the
lattice currents, $J_0^{(i)}$, at the vertex and then
project out the tree level expression $\langle J_0^{(j)}
\rangle_{\text{tree}}$. Thus, for example, $\zeta_{10}$ represents the
projection of $J_0^{(1)}$ onto $J_0^{(0)}$ and $\zeta_{11}$ the
projection of $J_0^{(1)}$ onto itself.

Some of the mixing matrix elements are infrared divergent and, as for the
wavefunction renormalisation contributions, we separate the infrared
divergent and finite pieces. For example, we write 
\begin{equation}
 \widetilde{\zeta}_{00} = \zeta_{00} + \zeta_{00}^{\text{IR}},
\end{equation}
where
\begin{equation}
\zeta_{00}^{\text{IR}} = -\frac{1}{3\pi}\log(a^2\lambda^2).
\end{equation}
We confirm that all infrared divergences ultimately cancel in the matching
coefficients $\rho_i$. Demonstrating that the matching coefficients are
infrared finite is a nontrivial check of our results.

The matrix element $\zeta_{02}$ includes a term that removes an ${\cal
O}(a\als)$ discretisation error from $J_0^{(0)}$
\cite{morningstar98,gulez04}. Thus the matching procedure ensures that 
${\cal O}(\als/M_0)$ and ${\cal O}(a\als)$ corrections are made at the same
time.

Finally we note that there is a second dimension four current operator
that is equivalent to $J_0^{(2)}$ via the equations of
motion \cite{morningstar98,gulez04}:
\begin{equation}
\widetilde{J}_0^{(2)} =
\frac{1}{2(aM_0)}\,\overline{q}(x)\frac{\overleftarrow{\partial}}{\partial t}\,
\Gamma_0 \,Q(x),
\end{equation}
where the arrow indicates that the derivative acts to the left. The effects of
this current operator must be included in the
determination of $\zeta_{i2}$.

\subsubsection{Spatial vector current}

In this case we require only the first two of the
three lattice currents given above, those of Equations \eqref{eq:j0def} and
\eqref{eq:j1def}. The matrix element of $V_k$ in full QCD is related to the
effective NRQCD current via
\begin{equation}
\langle V_k\rangle = \left(1+\als \rho_0^{\,(V_k)}\right)
\langle J_k^{(0)}\rangle + \left\langle
J_k^{(1),\,\text{sub}}\right\rangle,\label{eq:jvk}
\end{equation}
where
\begin{equation}
J_\mu^{(1),\,\text{sub}} = J_\mu^{(1)} - \als \zeta_{10}J_\mu^{(0)}
\end{equation}
and
\begin{align}
\rho_0^{\,(V_k)} = {} & B_0^{\,(V_k)} - \frac{1}{2}(C_q+C_H) -
\zeta_{00}^{\,(V_k)}, \label{eq:rho0vk} \\
B_0^{\,(V_k)} = {} & \frac{1}{\pi}\left(\ln(aM_0) - \frac{11}{12}\right).
\end{align}

The only contribution to $\zeta_{00}^{\,(V_k)}$
and
$\zeta_{10}^{\,(V_k)}$ is the vertex correction diagram in Figure
\ref{fig:zetaijfeyndiags} with the current $J_0^{(0)}$ or $J_0^{(1)}$ inserted
at the
vertex.

\subsection{Massive Quarks}

The matching calculation for massive HISQ quarks proceeds in a similar
manner to the massless case just discussed. Here, however, one must rescale the
lattice currents, $J_\mu^{(i)}$, by the tree-level massive HISQ wavefunction
renormalization $\left(Z_Q^{(0)}\right)^{-1/2}$. In the following we assume
that the currents have been rescaled.

\subsubsection{Vector current}

We again require only two of the three lattice currents:
$J_\mu^{(0)}$ and $J_\mu^{(1),\,\text{sub}}$. We write the
matrix element of the vector current in full QCD
in terms of the matrix elements of $J_0^{(0)}$ and $J_\mu^{(1),\,\text{sub}}$ as
\begin{align}
\langle V_\mu \rangle = \left(1+\als
\eta_0^{\,(V_\mu)}\right) \label{eq:vmu}
\langle J_\mu^{(0)}\rangle+ \left\langle J_\mu^{(1),\,\text{sub}}\right\rangle,
\end{align}
where, in this case,
\begin{equation}
J_\mu^{(1),\,\text{sub}} = J_\mu^{(1)} - \als \tau_{10}J_\mu^{(0)}.
\end{equation}
We denote the matching coefficient for massive HISQ quarks by
$\eta_0$, to clearly distinguish the massless and massive cases. The matching
coefficient is given
by
\begin{equation}
\eta_0^{\,(V_\mu)} = D_0^{\,(V_\mu)} - \frac{1}{2}(C_Q+C_H) -
\tau_{00}^{\,(V_\mu)},
\end{equation}
with \cite{kuramashi98,boyle00}
\begin{align}
D_0^{\,(V_0)} = {} & \frac{1}{\pi}\left(\frac{a\mtree+aM_0}{a\mtree-aM_0}\ln
\left(\frac{a\mtree}{aM_0}\right) - 2\right), \\
D_0^{\,(V_k)} = {} &
\frac{1}{\pi}\bigg(
\frac{a\mtree-aM_0}{a\mtree+aM_0}\ln \left(\frac{a\mtree}{aM_0} \right) -
\frac{8}{3}\bigg).
\end{align}
The $\tau_{ij}$ are the mixing
matrix elements for massive relativistic quarks.

The leading order mixing matrix elements for the temporal vector current are
logarithmically infrared divergent. Hence we write 
\begin{equation}
\widetilde{\tau}_{00}^{\,(V_0)} = \tau_{00}^{\,(V_0)} + \tau_{00}^{\text{IR}},
\end{equation}
where
\begin{equation}
\tau_{00}^{\text{IR}} = \frac{2}{3\pi}\log(a^2\lambda^2).
\end{equation}
In contrast to the massless case the infrared divergences in the
vertex and wavefunction renormalisations cancel separately in both the lattice
and continuum matrix elements. Confirming that the sum of the lattice
results is infrared finite serves as an
important cross-check of our calculation.

The evaluation of the mixing matrix elements for the spatial vector current is
more complicated
than for the temporal component. In this case the mixing matrix element
$\tau_{00}^{\,(V_k)}$
contains not only a logarithmic divergence but a linear divergence as well:
\begin{align}
\widetilde{\tau}_{00}^{\,(V_k)} = {} & \tau_{00}^{\,(V_k)} +
\tau_{00}^{\text{IR}}, \nonumber\\
\tau_{00}^{\text{IR}} = {} &
\frac{1}{3\pi}\left(8\pi \frac{a\mtree aM_0}{a\mtree+aM_0}\frac{1}{a\lambda}
+2\log(a^2\lambda^2)\right).\label{eq:tauir}
\end{align}
The logarithmic divergence is cancelled by the wavefunction renormalization,
leaving both
lattice and continuum contributions with a linear divergence. This, in turn,
cancels when we 
match the lattice and continuum results so that the matching
coefficient is infrared finite.

\subsubsection{Axial-vector current}

The matching relation for the axial-vector current is given at leading
order by
\begin{equation}
\langle A_\mu \rangle = \left(1+\als \eta_0^{\,(A_\mu)}\right)\label{eq:amu}
\langle J_\mu^{(0)}\rangle+ \left\langle J_\mu^{(1),\,\text{sub}}\right\rangle.
\end{equation}
Here we have
\begin{equation}
\eta_0^{\,(A_\mu)} = D_0^{\,(A_\mu)} - \frac{1}{2}(C_Q+C_H) -
\tau_{00}^{\,(A_\mu)},
\end{equation}
where \cite{kuramashi98}
\begin{align}
D_0^{\,(A_k)} = {} & \frac{1}{\pi}\left(
\frac{a\mtree+aM_0}{a\mtree-aM_0}\ln \left(\frac{a\mtree}{aM_0} \right) -
\frac{8}{3}\right), \\
D_0^{\,(A_0)} = {} & \frac{1}{\pi}\bigg(\frac{a\mtree-aM_0}{a\mtree+aM_0}\ln
\left(\frac{a\mtree}{aM_0} \right) -2
\bigg).
\end{align}
Here the $A_0$ current develops a linear IR divergence, which is the same as
that given in Equation \eqref{eq:tauir} for $V_k$. This divergence again cancels
between lattice and continuum results.

In the following section we present our results for the matching coefficients
$\rho_i$ of Equations \eqref{eq:jv0} and \eqref{eq:jvk} and $\eta_0$ of
\eqref{eq:vmu} and \eqref{eq:amu}, together 
with the mixing matrix elements $\zeta_{10}$ and $\tau_{10}$ needed to fix
$J_\mu^{(i),\,\text{sub}}$.

\section{\label{sec:matchresults}Matching procedure results}

As we discussed in the previous results section for the quark
renormalisation parameters, Section \ref{sec:zqresults}, we implement
two independent calculation procedures to cross-check our results. We
have calculated all the relevant mixing matrix elements, $\zeta_{ij}^{V_\mu}$
and $\tau_{ij}^{\Gamma_\mu}$, required to match the lattice currents with
continuum QCD. For clarity of presentation, however, we only give our results
for the final matching coefficients, $\rho_i^{\,V_\mu}$ and
$\eta_0^{\,\Gamma_\mu}$. We also include the mixing matrix elements,
$\zeta_{10}^{V_\mu}$ and $\tau_{10}^{\Gamma_\mu}$, for
completeness, because these are needed to construct the subtracted lattice
currents $J_\mu^{(i),\,\text{sub}}$.

\subsection{Massless Quarks}

\begin{figure}
\includegraphics[height=0.12\textwidth,width=0.2\textwidth]{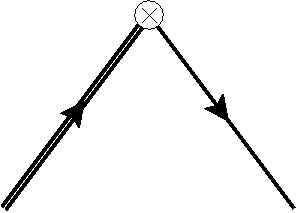}
\caption{\label{fig:tadcorrdiag}Tadpole correction contribution to the
one loop mixing matrix element $\rho_1$. The double lines indicate heavy quarks,
the single line the light quark and the cross represents the tadpole insertion.}
\end{figure}

We tabulate our results for the matching coefficients
$\rho_i^{\,(V_0)}$ at four
different heavy quark masses in Table \ref{tab:zetaijA0}.
\begin{table}
\caption{\label{tab:zetaijA0}One-loop matching for
the temporal vector current with NRQCD heavy quarks and massless HISQ
light 
quarks. All results use stability parameter
$n=4$ in the NRQCD action. We implement tadpole improvement with the Landau
link definition of $u_0$. For the $\rho_i^{\,(V_0)}$ the quoted
uncertainties
are the errors from each contribution added in quadrature, whilst for
$\zeta_{10}^{\,(V_0)}$ the uncertainty is purely the statistical error
from numerical integration. \\}
\begin{ruledtabular}
\begin{tabular}{ccccc}
\vspace*{-5pt}\\
$aM_0$ &  $\rho_0^{\,(V_0)}$ & $\rho_1^{\,(V_0)}$ &
$\rho_2^{\,(V_0)}$ & $\zeta_{10}^{\,(V_0)}$ \\
\vspace*{-5pt}\\
\hline
\vspace*{-5pt}\\
3.297 & -0.072(2) & \hphantom{-}0.048(2) & -1.108(4) & -0.0958(1) \\
3.263 & -0.075(2) & \hphantom{-}0.046(2) & -1.083(4) & -0.0966(1) \\
3.25\hphantom{0} & -0.075(1) & \hphantom{-}0.046(2) & -1.074(4) & -0.0970(1) \\
2.688 & -0.109(2) & \hphantom{-}0.013(2) & -0.712(4) & -0.1144(1) \\
2.66\hphantom{0} & -0.110(2) & \hphantom{-}0.013(2) & -0.698(4) & -0.1156(1) \\
2.650 & -0.112(2) & \hphantom{-}0.013(2) & -0.696(4) & -0.1157(1) \\
2.62\hphantom{0} & -0.116(2) & \hphantom{-}0.008(2) & -0.690(4) & -0.1171(1) \\
1.91\hphantom{0} & -0.161(2) & -0.038(3) & -0.325(4) & -0.1539(1) \\
1.89\hphantom{0} & -0.162(2) & -0.038(3) & -0.318(4) & -0.1553(1) \\
1.832 & -0.162(2) & -0.042(3) & -0.314(4) & -0.1593(2) \\
1.826 & -0.163(3) & -0.043(3) & -0.311(4) & -0.1595(2) \\
\vspace*{-5pt}\\
\end{tabular}
\end{ruledtabular}
\end{table}
Only the matching coefficient $\rho_1$ has a tadpole correction
coefficent, arising from the tadpole correction insertion illustrated in
Figure \ref{fig:tadcorrdiag}.
This correction contributes to $\zeta_{11}^{\,(V_0)}$ and is given by
\begin{equation}
\zeta_{11}^{u_0} = u_0^{(1)}.
\end{equation}
We use the Landau link definition of the tadpole correction coefficient,
$u_0^{(1)} = 0.7503(1)$, when calculating $\rho_1$.

In Table \ref{tab:zetaijVk} we give our results for the matching
coefficients for the spatial components of the heavy-light vector current
with massless HISQ light quarks and NRQCD heavy quarks.
\begin{table}
\caption{\label{tab:zetaijVk}One-loop matching coefficients for
the spatial vector current with massless HISQ
light quarks. See the caption accompanying Table \ref{tab:zetaijA0} for more details.\\}
\begin{ruledtabular}
\begin{tabular}{ccc}
\vspace*{-5pt}\\
$aM_0$ &  $\rho_0^{\,(V_k)}$ & $\zeta_{10}^{\,(V_k)}$ \\
\vspace*{-5pt}\\
\hline
\vspace*{-5pt}\\
3.297 & -0.046(2) & 0.0319(1) \\
3.263 & -0.045(2) & 0.0322(1) \\
3.25\hphantom{0} & -0.045(2) & 0.0323(1) \\
2.688 & -0.034(2) & 0.0382(1) \\
2.66\hphantom{0} & -0.034(2) & 0.0385(1) \\
2.650 & -0.034(2) & 0.0386(1) \\
2.62\hphantom{0} & -0.033(2) & 0.0391(1) \\
1.91\hphantom{0} & 0.007(2) & 0.0513(1) \\
1.89\hphantom{0} & 0.009(2) & 0.0518(1) \\
1.832 & 0.020(2) & 0.0532(1) \\
1.826 & 0.020(2) & 0.0532(1) \\
\vspace*{-5pt}\\
\end{tabular}
\end{ruledtabular}
\end{table}

\subsection{Massive Quarks}

In Table \ref{tab:tauijV0scatter} we tabulate our results for the matching
coefficients for $V_0$
with massive HISQ light quarks and NRQCD heavy quarks.
\begin{table}
\caption{\label{tab:tauijV0scatter}One-loop matching coefficients for the
temporal vector current with massive HISQ
quarks. All results use stability parameter
$n=4$ in the NRQCD action. For $\eta_0^{\,(V_0)}$ the quoted
uncertainties
are the errors from each contribution added in quadrature, whilst for
$\tau_{10}^{\,(V_0)}$ the uncertainty is purely the statistical error
from numerical integration. \\}
\begin{ruledtabular}
\begin{tabular}{cccc}
\vspace*{-5pt}\\ 
$aM_0$ & $am_0$ & $\eta_0^{\,(V_0)}$ & $\tau_{10}^{\,(V_0)}$ \\
\vspace*{-5pt}\\
\hline
\vspace*{-5pt}\\
3.297 & 0.8260 & -0.151(3) & -0.0488(1) \\ 
3.263 & 0.8180 & -0.148(3) & -0.0494(1) \\ 
2.688 & 0.6300 & -0.121(3) & -0.0647(1) \\
2.660 & 0.6450 & -0.117(3) & -0.0648(1) \\
2.650 & 0.6235 & -0.113(3) & -0.0658(1) \\
2.650 & 0.6207 & -0.112(3) & -0.0659(1) \\
2.620 & 0.6270 & -0.116(3) & -0.0663(1) \\
1.910 & 0.4340 & -0.102(3) & -0.0990(1) \\
1.832 & 0.4130 & -0.098(3) & -0.1043(1) \\
1.826 & 0.4120 & -0.098(3) & -0.1046(1) \\
\vspace*{-5pt}\\
\end{tabular}
\end{ruledtabular}
\end{table}

We give our results for the matching coefficients for $V_k$
in Table \ref{tab:tauijVkannihil}. Finally, in Tables
\ref{tab:tauijA0} and \ref{tab:tauijAk}, we present our results for the matching coefficients
for $A_0$ and $A_k$ respectively.

\begin{table}
\caption{\label{tab:tauijVkannihil}One-loop matching for
spatial vector current with massive HISQ quarks. See the caption accompanying
Table \ref{tab:tauijV0scatter} for
more details. \\}
\begin{ruledtabular}
\begin{tabular}{cccc}
\vspace*{-5pt}\\
$aM_0$ & $am_0$ & $\eta_0^{\,(V_k)}$ & $\tau_{10}^{\,(V_k)}$ \\
\vspace*{-5pt}\\
\hline
\vspace*{-5pt}\\
3.297 & 0.8260 & -0.124(5) & 0.0420(1) \\ 
3.263 & 0.8180 & -0.118(5) & 0.0423(1) \\ 
2.688 & 0.6300 & -0.025(5) & 0.0484(1) \\
2.660 & 0.6450 & -0.024(5) & 0.0488(1) \\
2.650 & 0.6235 & -0.015(5) & 0.0489(1) \\
2.650 & 0.6207 & -0.014(5) & 0.0489(1) \\
2.620 & 0.6270 & -0.019(5) & 0.0493(1) \\
1.910 & 0.4340 & \hphantom{-}0.049(5) & 0.0618(1) \\
1.832 & 0.4130 & \hphantom{-}0.059(5) & 0.0636(1)\\
1.826 & 0.4120 & \hphantom{-}0.060(5) & 0.0638(1) \\
\vspace*{-5pt}\\
\end{tabular}
\end{ruledtabular}
\end{table}

\begin{table}
\caption{\label{tab:tauijA0}One-loop matching for
the temporal axial-vector current with massive HISQ quarks. See the caption
accompanying Table \ref{tab:tauijV0scatter} for
more details. \\}
\begin{ruledtabular}
\begin{tabular}{cccc}
\vspace*{-5pt}\\
$aM_0$ & $am_0$ & $\eta_0^{\,(A_0)}$ & $\tau_{10}^{\,(A_0)}$ \\
\vspace*{-5pt}\\
\hline
\vspace*{-5pt}\\
3.297 & 0.8260 & -0.237(5) & -0.1260(1) \\ 
3.263 & 0.8180 & -0.232(5) & -0.1269(1) \\ 
2.688 & 0.6300 & -0.188(5) & -0.1452(1) \\
2.660 & 0.6450 & -0.192(5) & -0.1464(1) \\
2.650 & 0.6235 & -0.183(5) & -0.1468(1) \\
2.650 & 0.6207 & -0.182(5) & -0.1467(1) \\
2.620 & 0.6270 & -0.189(5) & -0.1480(1) \\
1.910 & 0.4340 & -0.219(5) & -0.1853(1) \\
1.832 & 0.4130 & -0.222(5) & -0.1908(1) \\
1.826 & 0.4120 & -0.221(5) & -0.1914(1) \\
\vspace*{-5pt}\\
\end{tabular}
\end{ruledtabular}
\end{table}

\begin{table}
\caption{\label{tab:tauijAk}One-loop matching for
the spatial axial-vector current with massive HISQ quarks. See the caption
accompanying Table \ref{tab:tauijV0scatter} for
more details. \\}
\begin{ruledtabular}
\begin{tabular}{cccc}
\vspace*{-5pt}\\
$aM_0$ & $am_0$ & $\eta_0^{\,(A_k)}$ & $\tau_{10}^{\,(A_k)}$ \\
\vspace*{-5pt}\\
\hline
\vspace*{-5pt}\\
3.297 & 0.8260 & -0.260(3) & 0.0163(1) \\ 
3.263 & 0.8180 & -0.260(3) & 0.0165(1) \\ 
2.688 & 0.6300 & -0.194(3) & 0.0216(1) \\
2.660 & 0.6450 & -0.191(3) & 0.0216(1) \\
2.650 & 0.6235 & -0.183(3) & 0.0219(1) \\
2.650 & 0.6207 & -0.182(3) & 0.0320(1) \\
2.620 & 0.6270 & -0.185(3) & 0.0221(1) \\
1.910 & 0.4340 & -0.091(3) & 0.0330(1) \\
1.832 & 0.4130 & -0.076(3) & 0.0348(1) \\
1.826 & 0.4120 & -0.076(3) & 0.0349(1) \\
\vspace*{-5pt}\\
\end{tabular}
\end{ruledtabular}
\end{table}

\section{\label{sec:summary}Summary}

We have calculated the one loop matching coefficients required to match
the axial-vector and vector currents on the lattice to full QCD. We
used the HISQ action, with both massless and massive quarks, for the light
quarks and NRQCD for the heavy quarks. As part of the matching procedure
we have presented one loop mass and wavefunction renormalisations for both
HISQ and NRQCD quarks. We find that the perturbative coefficients are well
behaved and none are unduly large.

The matching coefficients for HISQ-NRQCD currents with massless HISQ quarks are
important ingredients in the determination
of heavy-light mesonic decay parameters from lattice QCD studies \cite{na12}.
Recent
studies of the $B_s$ meson using the relativistic HISQ action for both $b$
and $s$ quarks have been carried out \cite{mcneile12a}. Such an
approach has the advantage that perturbative matching, which
is generally the dominant source of error in the extraction of decay
constants, is not required. Currently, however, simulations at the
physical $b$ quark mass are prohibitively expensive and an extrapolation
up to the $b$ quark mass is still needed. Furthermore, simulations of the
$B$ meson are not presently feasible, as the use of light valence quarks
and close-to-physical $b$ quark masses requires both large lattices and
fine lattice spacings. In light of these considerations, the use of an
effective theory for heavy-light systems remains the most efficient method
for precise predictions of $f_{B_s}/f_B$ and $f_B$. Such calculations
require the perturbative matching calculation reported in this article. 

The matching calculations reported in this work are also crucial for the
HPQCD collaboration's nonperturbative studies of
the semileptonic decays of $B$ and $B_s$ mesons with NRQCD and HISQ
quarks. On the one hand, matching coefficients with massless HISQ quarks are
required for the determination of the
$B\rightarrow \pi\ell\nu$, $B\rightarrow K\ell^+\ell^-$ and $B_s\rightarrow
K\ell\nu$ decay parameters \cite{bouchard12}. On the other hand, our results for
the matching coefficients with massive HISQ quarks will be needed in future
calculations of the $B\rightarrow D\ell\nu$ and $B_s\rightarrow D_s\ell\nu$
decay parameters.

\begin{acknowledgments}
The authors would like to thank Georg von Hippel and Peter Lepage for many helpful
discussions during the course of this work. This work was supported by the
U.S. DOE, Grants No.~DE-FG02-04ER41302 and DE-FG02-91ER40690. Some of the
computing was undertaken on the Darwin supercomputer at the HPCS, University of
Cambridge, as part of the DiRAC facility jointly funded by the STFC.
\end{acknowledgments}

\appendix

\section{\label{app:oneloop}The HISQ tuning parameter}

In this appendix, we derive expressions for the tree level and one loop
tuning parameters, $\epsilon_{\text{tree}}$ and $\epsilon_1$. Throughout this
appendix we neglect factors of the lattice spacing, $a$, for convenience.

For an onshell particle with momentum given by $p_{\mu} = (iE,0,0,p_z)$ one
defines
the kinetic mass as
\begin{equation}
\mkin = \left(\frac{\partial^2 E}{\partial p_z^2}\right)^{-1}_{p_z = 0}.
\end{equation}
At nonzero momentum the tree level pole condition becomes
\begin{align}
m_0^2  = {} & \left[\sinh(E) -
\frac{1+\epstree}{6}\left(\sinh(E)\right)^3\right]^2 \nonumber \\
{} & - \left[\sin(p_z) +
\frac{1+\epstree}{6}\left(\sin(p_z)\right)^3\right]^2,
\end{align}
which, for notational convenience, we write as
\begin{equation}
m_0^2 = [X(E)]^2-[Y(p_z)]^2.
\end{equation}

Using the relations 
\begin{equation}
Y(p_z=0) = 0,\quad \frac{\partial E}{\partial p_z}\bigg|_{p_z = 0} =
0,\quad
\frac{\partial Y(p_z)}{\partial p_z}\bigg|_{p_z = 0} = 1,
\end{equation}
and differentiating twice using
\begin{equation}\label{eq:diffrelation}
\frac{\text{d}}{\text{d}p_z} = \frac{\partial}{\partial p_z} +
\frac{\partial E}{\partial p_z}\frac{\partial}{\partial E},
\end{equation}
we find
\begin{equation}
\left[\left(-X\frac{\partial X}{\partial E}\right)\left(\frac{\partial^2
E}{\partial p_z^2}\right) + \left(\frac{\partial Y}{\partial
p_x}\right)^2\right]_{p_z = 0} = 0,
\end{equation}
and thus the tree level kinetic mass is
\begin{align}
\mkin^{(0)} = {} & X\frac{\partial X}{\partial E} \nonumber \\
= {} &
\cosh(\mtree)\sinh(\mtree)(1-\Theta)(1-3\Theta). \label{eq:mkintree}
\end{align}
Here we have defined $\Theta = (1+\epstree)(\sinh(\mtree))^2/6$.

Requiring $\mkin^{(0)}=\mtree$ imposes a condition on the tree level
tuning parameter that leads to Equation \eqref{eq:epstreedef}.

At one loop, the procedure is much the same. The one loop
pole condition is
\begin{align}
(m_0-\alpha_s\sigmaB)^2 = {} & \left[\widetilde{X}(E) -
\sinh(E)\alpha_s\sigmaA_0\right]^2 \nonumber \\
{} & - \left[Y(p_z)-\sin(p_z)\alpha_s\sigmaA_z\right]^2,
\end{align}
where
\begin{equation}
\widetilde{X}(E) = \sinh(E)\left[1 -
\frac{1+\epsilon}{6}\left(\sinh(E)\right)^2\right].
\end{equation}

\begin{widetext}
Differentiating twice using Equation \eqref{eq:diffrelation} leads, after
some algebra, to
\begin{align}
\mkin = {} & \widetilde{X}\frac{\partial \widetilde{X}}{\partial E} -
\alpha_s\bigg\{\sigmaA_0\left(Z_Q^{(0)}\right)^{-1}\sinh(\mtree) 
 + 2\mtree
\sigmaA_z -
m_0\frac{\partial}{\partial E}\left[\sinh(E)\sigmaA_0 -
\sigmaB\right]
\nonumber \\
{} & - m_0\mtree
\frac{\partial^2}{\partial p_z^2}\left[\sinh(\mtree)\sigmaA_0
-\sigmaB\right]\bigg\},
\end{align}
where we have only kept terms up to ${\cal O}(\alpha_s)$.

For convenience, we write this as
\begin{equation}\label{eq:mkin}
\mkin = \widetilde{X}\frac{\partial \widetilde{X}}{\partial E} + \alpha_s
\sigma.
\end{equation}
Using the expansions of Equations \eqref{eq:mexpansion},
\eqref{eq:epsilonexpansion} and \eqref{eq:m1}, we 
can evaluate the product, $\widetilde{X}\frac{\partial
\widetilde{X}}{\partial E}$, at one loop to obtain
\begin{equation}\label{eq:xdx}
\widetilde{X}\frac{\partial \widetilde{X}}{\partial
E}=\mtree+u_1\epsilon_1\alpha_s
+u_2m_1\alpha_s,
\end{equation}
where
\begin{align}
u_1 = {} & \frac{1}{6}\cosh(\mtree)(\sinh(\mtree))^3\Big((1+\epstree)
(\sinh(\mtree))^2-4\Big)
\\
u_2 = {} &(\sinh(\mtree))^2\Big(1-\frac{1+\epstree}{6}
(\sinh(\mtree))^2\Big)\bigg[1-\frac{1+\epstree}{2}
\Big(2(\cosh(\mtree))^2+(\sinh(\mtree))^2\Big)\bigg]
 \nonumber \\
{} & +(\cosh(\mtree))^2\Big(1-\frac{1+\epstree}{2}
(\sinh(\mtree))^2\Big)^2
\end{align}
\end{widetext}

We can therefore write Equations \eqref{eq:mkin} and \eqref{eq:xdx} as
\begin{equation}
\mkin = \mtree + \alpha_s\Big(\nu^{(\epsilon)}\epsilon_1 +
\nu^{(\Sigma)}\Big),
\end{equation}
with $\nu^{(\Sigma)}$ and $\nu^{(\epsilon)}$ given by
\begin{align}
\nu^{(\Sigma)} = {} & u_2m_1^{(\Sigma)} + \sigma, \label{eq:nu1}\\
\nu^{(\epsilon)} = {} & u_1+u_2m_1^{(\epsilon)}, \label{eq:nueps}
\end{align}
and
\begin{align}
m_1^{(\Sigma)} = {} & \sinh(\mtree)\sigmaA_0 - \sigmaB, \\
m_1^{(\epsilon)} = {} & \frac{1}{6}Z_Q^{(0)}(\sinh(\mtree))^3.
\end{align}

We obtain an expression for the one loop tuning parameter by equating
the one loop masses: $\mkin^{(1)}
= m_1^{(1)}$. The result is
\begin{equation}
\epsilon_1 =
\frac{m_1^{(\Sigma)}-\nu^{(\Sigma)}}{\nu^{(\epsilon)}-m_1^{(\epsilon)}}
\end{equation}
where $\nu^{(\Sigma)}$ and $\nu^{(\epsilon)}$ are given in Equations
\eqref{eq:nu1} and \eqref{eq:nueps} respectively.

\section{\label{app:subfn} Subtraction functions for numerical integration}
At intermediate stages of the lattice-to-continuum matching procedure 
one encounters infrared (IR) divergent integrals and care is
required to ensure that \vegas can handle them accurately. For diagrams involving massless HISQ
fermions, it is usually sufficient 
to introduce a nonzero gluon mass $\lambda$, fit results to appropriate 
functions of this mass and then extract the IR finite parts. For massive HISQ fermions, on the
other hand, it is often necessary to include specific 
 subtraction terms into the integrand  in order to 
stabilize the \vegas integrations. 
In this appendix we list examples of such subtraction terms. Given an IR divergent integral,
\begin{equation}
\label{fint}
\cali = C_F\intk  \,\calf_{\text{lat}}(k,\lambda),
\end{equation}
where
$C_F = 4/3$ is a color factor (the quadratic Casimir operator) required to
correctly
normalize the infrared divergences. We employ subtraction terms in the following
way:
\begin{align}
\cali = {}& C_F\intk \left \{ \calf_{\text{lat}}(k,\lambda)  -
\calf_{\text{sub}}
(k,m_{eff},\Lambda,\lambda) \right \} \nonumber \\
  &{} \qquad  + F(\meff,\Lambda,\lambda),\label{eq:subtract}
\end{align}
where
\begin{equation}
 F(\meff,\Lambda,\lambda) =
 C_F\intk  {\cal F}_{\text{sub}}(k,m_{\it
eff},\Lambda,\lambda).\label{eq:addback}
\end{equation}
Here $\Lambda$ is a cutoff imposed on $\calf_{\text{sub}}$ such that
$\calf_{\text{sub}}
\equiv 
0$ for $k^2 \geq \Lambda^2$, and $\meff$ is defined below. The full expression 
for $\cali$ in \eqref{eq:subtract} does not, of course, depend on $\Lambda$. We
have done the
calculations with two different values for $\Lambda$, 
e.g. $a \Lambda$ = 2 and 3, to check this.
 
The choice for $\calf_{\text{sub}}$ is far from unique. One wants a function of 
$k_\mu$ with the same IR behaviour as the original integrand
$\calf_{\text{lat}}$, 
that is, however, simple enough that the integral in the ``addback'' function,
$F(\meff,\Lambda,\lambda) $, can be evaluated with relative ease. One
natural choice 
is the integrand of the corresponding continuum theory Feynman diagram
$\calf_{\text{cont}}$.  
This is what has often been done in the literature. For massive fermions 
there remains the question  of what fermion mass to use in
$\calf_{\text{cont}}$.  
It was suggested in \cite{kuramashi98} to pick a mass, denoted by $\meff$, such
that 
$\calf_{\text{sub}}(k,\meff,\Lambda,\lambda)$ mimics as closely as possible 
the correct $k_\mu \rightarrow 0$ limit in the denominator 
of the lattice fermion propagator. For instance in the continuum theory 
one would have (we work in Euclidean space), for onshell quarks with 
external momentum $p_\mu \rightarrow (im,\vec{0})$, a fermion propagator with
denominator given
by
\begin{equation}
\text{denom} = (p-k)^2 + m^2 \rightarrow k^2 -2 i m k_0.\label{eq:contdenom}
\end{equation}
Taking a hint from \eqref{eq:contdenom} we pick $\meff$ by first setting the
external momentum to
$p_\mu = (i \mtree, \vec{0})$, expanding 
the denominator of the free HISQ propagator around $k_0 = 0$ and then looking
for the coefficient
of $ (-2 i k_0)$. One finds,
\begin{equation}
\label{eq:meff}
\meff = \cosh(\mtree)\sinh(\mtree)  
(1 - \Theta) (1 - 3 \Theta),
\end{equation}
where $\Theta$ is defined after \eqref{eq:mkintree}. We recognize this as
$\mkin^{(0)}$, given in \eqref{eq:mkintree},
so that
\begin{equation}
\label{eq:meff2}
\meff = \mkin^{(0)} \equiv \mtree,
\end{equation}
a result that may not come as a surprise. We note, however, that the last
equality 
in \eqref{eq:meff2} holds only because we have tuned $\epstree$ to
ensure 
$\mkin^{(0)} = \mtree = m^{(0)}_{\text{pole}}$.  This was not the case for
uses of 
$\meff$ in the past \cite{kuramashi98,groote00} involving 
massive Clover fermions.

Following the guidlines described above, the subtraction term for the rainbow 
diagram correction to the massive HISQ wave function renormalization $Z_Q$ 
becomes
\begin{eqnarray}
\calf_{\text{sub}}^{Z_Q} & = & \theta(\Lambda^2 - k^2) \left \{ \frac{4 (k_0^2 
+ b^2/4) \,((k^2)^2 - b^2 k_0^2)} {(k^2 + \lambda^2) \, ((k^2)^2 + b^2
k_0^2)^2} 
\right. \nl
&& - ( \xi -1) \left. \frac{k_0^2 (b^2 + 2 k^2)}{k^2 (k^2 + \lambda^2)\,
((k^2)^2 + b^2 k_0^2)} \right \}, 
\end{eqnarray}
with $b = 2 \mtree$. This leads to an addback function
\begin{widetext}
\begin{align}
F^{Z_Q} = {} &  
\frac{2 }{3 \pi} \bigg\{  \left [\log \left (\frac{\Lambda^2}{\lambda^2}
\right ) 
 - \log \left ( \frac{W_1}{b} \right )  -2
\frac{ \Lambda^2}{b^4} (b^2 + 3 \Lambda^2) 
 - \frac{\Lambda W_0}{b^4}  (b^2 - 6 \Lambda^2) \right ] \nonumber \\
{} &  - \frac{(\xi - 1)}{2} \left [ \log \left (\frac{\Lambda^2}{\lambda^2}
\right
) - \log \left ( \frac{W_1}{b} \right ) 
 + 2\frac{ \Lambda^2}{b^4} (2 b^2+\Lambda^2  )
-\frac{\Lambda  W_0}{b^4} (3 b^2+2 \Lambda^2)
\right ] \bigg\},
\end{align}
where $W_0 \equiv \sqrt{b^2 + \Lambda^2}$ and $W_1 \equiv [\Lambda + W_0]$.
\end{widetext}

Similarly, for the one-loop vertex correction for a scattering diagram
involving 
$V_0$ one has in Feynman gauge,
\begin{equation}
\label{eq:v0}
\calf_{\text{sub}}^{V_0} = \frac{ \theta(\Lambda^2 - k^2) 
 \left \{k_0^2 \, (k^2 + b^2) \; - \; 
\frac{b}{2 M} (\vec{k}^2)^2 \right \}}
{[(k^2)^2 + b^2 k_0^2] \, [k_0^2 + (\vec{k}^2/(2M))^2] \, [k^2 + \lambda^2]}.
\end{equation}
And for the annihilation diagram involving $V_k$ one has,
\begin{equation}
\label{eq:vk}
\calf_{\text{sub}}^{V_k} = \frac{ \theta(\Lambda^2 - k^2) \left \{ k_0^2 \, (k^2
+ b^2) \; +
\; 
\frac{b}{2 M} (\vec{k}^2)^2 \right \} }
{[(k^2)^2 + b^2 k_0^2] \, [k_0^2 + (\vec{k}^2/(2M))^2] \, [k^2 + \lambda^2]}.
\end{equation}
The only difference between \eqref{eq:v0} and \eqref{eq:vk} is the relative 
sign between the two terms in the numerator, i.e. the sign of the term linear 
in $\mtree$. This is as it should be, since for annihilation one has an 
incoming anti-HISQ quark with the on-shell condition $p_\mu 
\rightarrow (- i \mtree, \vec{0})$ replacing the outgoing HISQ quark of the
scattering 
process. The two terms in the numerator each lead to linear IR divergent
results 
which cancel in the case of $V_0$ leaving just a logarithmic IR divergence. 
For $V_k$ one ends up with an expression with both linear and logarithmic 
IR divergent terms as is required. We have not attempted
 to integrate $\calf_{\text{sub}}^{V_0}$ 
or $\calf_{\text{sub}}^{V_k}$ in closed form to obtain analytic expressions 
for the addback functions $F^{V_0}$ and $F^{V_k}$.  Instead we reduced the integrals to 1D 
integrals in the radial variable $0 \leq k \leq \Lambda$ and used \vegas again 
to evaluate them.



%


\end{document}